\documentclass[aps,prb,superscriptaddress,amsmath,amssymb,amsfonts]{revtex4}

\usepackage{bm}
\usepackage{graphicx}
\usepackage{color}

\usepackage{epsf}

\def\be{\begin{equation}}
\def\ee{\end{equation}}

\def\Tr{\mathop{\rm Tr}\nolimits}
\def\TrK{\mathop{\rm tr}} 
\def\tr{\mathop{\rm tr}}
\def\str{\mathop{\rm str}}

\def\diag{\mathop{\rm diag}}
\def\Re{\mathop{\rm Re}}
\def\Im{\mathop{\rm Im}}
\def\mls{\Delta}

\newcommand{\corr}[1]{\langle #1\rangle}

\def\bbar{{\bar b}}

\def\llangle{{\langle\!\langle}}
\def\rrangle{{\rangle\!\rangle}}

\def\eps{\varepsilon}

\begin{document}

\title{Dyson--Maleev representation of nonlinear sigma-models}

\author{D.~A.~Ivanov}
\affiliation{Institute for Theoretical Physics,
Ecole Polytechnique Federale de Lausanne (EPFL),
CH-1015 Lausanne, Switzerland}

\author{M.~A.~Skvortsov}
\affiliation{Landau Institute for Theoretical Physics, Chernogolovka,
Moscow region, 142432 Russia}

\date{January 14, 2008}

\begin{abstract}
For nonlinear sigma-models in the unitary symmetry class,
the non-linear target space can be parameterized with cubic
polynomials. This choice of coordinates has been known previously
as the Dyson-Maleev parameterization for spin systems, and
we show that it can be applied to a wide range of sigma-models.
The practical use of this parameterization includes simplification
of diagrammatic calculations (in perturbative methods) and of
algebraic manipulations (in non-perturbative approaches). We
illustrate the use and specific issues of the Dyson-Maleev
parameterization with three examples: the Keldysh sigma-model
for time-dependent random Hamiltonians, the supersymmetric
sigma-model for random matrices, and the supersymmetric
transfer-matrix technique for quasi-one-dimensional disordered wires.
We demonstrate that nonlinear sigma-models of unitary-like
symmetry classes C and B/D also admit the Dyson-Maleev parameterization.
\end{abstract}

\maketitle

\section{Introduction}

\subsection{Nonlinear sigma-models}

Extensive studies of disordered systems in last decades have
identified the nonlinear sigma-model (NLSM) formalism as a universal
tool for describing the low-energy properties of such systems
\cite{Wegner,Efetov1983,VWZ,Efetov-book,Mirlin2000,Finkelstein,BK1994,HorbachSchoen1993,KamenevAndreev99}.
The underlying idea of the NLSM approach is that the low-energy physics
of disordered systems is determined by soft collective excitations
(usually referred to as diffusons and cooperons). Therefore one can
integrate out the high-energy degrees of freedom and end up with
an effective theory which contains only soft collective modes. The
resulting theory is a NLSM, which is a field theory formulated in terms
of a matrix field $Q$ subject to a nonlinear constraint $Q^2=1$.
Depending on the way of handling disorder averaging, the $Q$-matrix
can act either in the supersymmetric \cite{Efetov1983,VWZ,Efetov-book,Mirlin2000},
replica \cite{Finkelstein,BK1994}
or Keldysh \cite{HorbachSchoen1993,KamenevAndreev99} spaces.
Besides that, the $Q$-field can be a function of continuous space
and/or time coordinates, and also have an additional matrix structure
due to an additional symmetry of the initial Hamiltonian.

The NLSM action usually has a hydrodynamic form, containing
only the lowest powers of $Q$, e.g.,
\be
  S[Q] = \Tr \left( [A,Q]^2+ B Q \right) ,
\label{S[Q]-general}
\ee
where $\Tr$ implies the full trace, involving
integration over continuous coordinates (in the supersymmetric formalism,
the trace over the superspace is the supertrace).
The action (\ref{S[Q]-general}) determines the weight $e^{-S[Q]}$ in
the functional integral which should be performed
over the manifold
\be
  Q=U^{-1}\Lambda U ,
\label{orbit}
\ee
where
\be
\Lambda=
\begin{pmatrix}
1 && 0 \\
0 && -1
\end{pmatrix}_{RA}
\label{Lambda-definition}
\ee
in the ``retarded-advanced'' space, and $U$ span an appropriate
group of rotations. The matrices (or operators)
$A$ and $B$ in (\ref{S[Q]-general}) must commute
with $\Lambda$: $[A,\Lambda]=0=[B,\Lambda]$,
so that $Q=\Lambda$ is a saddle point of the action.

Depending on the symmetries of the disordered problem, the initial
Hamiltonian may have additional symmetries, which translate into
additional linear constraints on the matrix $Q$ (realized in terms of
extending its dimension, in combination with imposing constraints on the
rotations $U$) \cite{Efetov-book,Zirnbauer-classification}.

The NLSM is a complicated field theory, and its exact solution is possible
only in a few exceptional cases: zero-dimensional supersymmetric sigma-model
for level statistics of random matrices\cite{Efetov1983,Efetov-book},
and one-dimensional supersymmetric sigma-model for quasi-one-dimensional
localization\cite{EL83,Efetov-book,Mirlin2000}.
In a situation where the integration over the whole manifold of $Q$ matrices
cannot be performed exactly, the sigma-model can be treated with the standard
perturbative approach, applicable in the weak-coupling limit
(large conductance).
Solution of the NLSM in the strong-coupling limit (small conductance)
is a challenging and so far unresolved task.

In the weak-coupling limit, a widely used perturbative expansion of the NLSM
around the saddle point $\Lambda$ is based on the parameterization of the curved
$Q$-matrix manifold in terms of an unconstrained matrix $W$:
\be
\label{Q-W}
  Q = \Lambda f(W),
\qquad
  W =
  \begin{pmatrix}
    0 & b \\
    -\bar b & 0
  \end{pmatrix}.
\ee
where $f(x)=1+x+x^2/2+c_3x^3+(c_3-1/8)x^4+\dots$ can be an arbitrary function
satisfying $f(x)f(-x)=1$.
A particular choice of the function $f(x)$ is a matter of convenience
(see discussion of various parameterizations in Ref.~\onlinecite{IS2006}), and
the calculated correlation functions do not depend on it.
The integration contour in the space of $b$ and $\bbar$ should be chosen consistently
with the convergence of the integrals performed in the derivation of the NLSM.
In the supersymmetric formalism, this amounts to the requirement of the compact
fermionic and non-compact bosonic sectors\cite{Efetov1983,VWZ}.

If the action (\ref{S[Q]-general}) possesses several saddle points,
an analogous perturbative expansion should be carried out around each of them.

Substituting $Q$ expressed in terms of $W$ into the initial action
(\ref{S[Q]-general}), one arrives at a new theory
\be
  \int \dots e^{-S[Q]} \, DQ
  =
  \int \dots e^{-S_f[W]} J_f[W] \, DW ,
\ee
with the action $S_f[W] = S[\Lambda f(W)]$ and, generally speaking,
with a Jacobian $J_f[W]$.
This new theory can be treated with the help of the standard perturbative
approach by separating the action into the Gaussian part, $S_f^{(2)}[W]$,
and the rest taken into account perturbatively.
However, the initial action (\ref{S[Q]-general}), which was a finite polynomial
in terms of $Q$, typically  becomes an infinite series in terms of $W$.
This makes higher-order perturbative calculations very involved
as one has to take into account a rapidly growing number of diagrams made of
various higher-order vertices $\propto W^n$.

\subsection{Dyson--Maleev parameterization}

For nonlinear sigma-models in the unitary symmetry class,
there exists a remarkable parameterization which does not belong
to the infinite-series family (\ref{Q-W}).
The non-linear target space can be parameterized by cubic polynomials:
\be
  Q =
  \begin{pmatrix}
    1-b\bar b/2 && b-b\bar bb/4 \\
    \bar b && -1+\bar bb/2
  \end{pmatrix}.
\label{Q-DM}
\ee
The asymmetry between the $b$ and $\bar b$ matrices is the price one has
to pay for finite number of terms in the parameterization (\ref{Q-DM}).
It can be shown by a direct calculation that the Jacobian of this transformation
is unity, and thus the integration measure over $b$ and $\bar b$
is flat: $DQ=Db\,D\bar b$. In the new variables, the NLSM action becomes
a finite-order polynomial, e.g., the action (\ref{S[Q]-general}) becomes
a sum of the bilinear and quartic in $b$, $\bar b$ terms:
\be
\label{S[Q]-DM}
  S[Q] = S^{(2)}[b,\bar b] + S^{(4)}[b,\bar b] .
\ee
The absence of higher-order interaction vertices greatly reduces the number
of diagrams in the perturbative expansion, which considerably simplifies
routine calculations.

The parameterization (\ref{Q-DM}) is closely related to
the famous Dyson-Maleev \cite{Dyson56,Maleev57} (DM) parameterization
for quantum spins. In that representation, the spin-$S$ operators
are expressed by the boson creation and annihilation
operators $\hat a^\dagger$ and $\hat a$ as
\be
  \hat S^+ = (2S-\hat a^\dagger \hat a) \hat a,
\qquad
  \hat S^- = \hat a^\dagger,
\qquad
  \hat S^z = S-\hat a^\dagger \hat a .
\label{DM-spin}
\ee
The Dyson-Maleev parameterization produces the correct spin
commutation relations but violates the property
$(\hat S^-)^\dagger=\hat S^+$, and thus
renders the spin Hamiltonian manifestly non-Hermitian.
An alternative approach based on the Holstein-Primakoff
parameterization \cite{HolstPrim40}
[which is in fact a counterpart of the parameterization (\ref{Q-W})
with $f(W)=\sqrt{1+W^2}+W$]
respects Hermiticity
but generates an infinite series of interaction vertices.
The Dyson--Maleev transformation has proven to be the most convenient tool
for studying spin-wave interaction in ferromagnets \cite{Canali92,Hamer92-93}:
it reproduces all the
perturbative results obtained with the Holstein-Primakoff
parameterization in a much faster and compact way.
Owing to the analogy with the Dyson--Maleev representation of
spin operators, the transformation (\ref{Q-DM}) will be referred
as the Dyson--Maleev representation of the NLSM.

In the original DM representation for spin operators,
the Hilbert space of free bosons should be truncated
in order for the operator $\hat S^z$ to have a bounded spectrum.
This truncation of the bosonic Hilbert space is irrelevant for
perturbative calculation but becomes essential in the non-perturbative
regime when the expectation value $\corr{\hat a^\dagger \hat a}$
is comparable to $S$.
Clearly, some analogy of this Hilbert space truncation should be present
also in the functional NLSM language. Indeed, in this paper we demonstrate
that imposing certain conditions on the eigenvalues of the matrix $b\bar b$
makes the DM representation non-perturbatively equivalent to the original
NLSM in terms of the $Q$ field.

At the {\em perturbative} level, the DM transformation (\ref{Q-DM})
has been applied for the analysis of the replicated\cite{Gruzberg1997}
and Keldysh\cite{IS2006} sigma-models
(cf.\ also the usage of the DM transformation \cite{Kolokolov2000}
for two-dimensional classical ferromagnets).
At this level, when fluctuations are small, it suffices to use
formally the parameterization (\ref{Q-DM}) without taking care
of the exact integration region over $b$ and $\bar b$.
We illustrate the details of the perturbation theory in the DM
parameterization with the examples of the Keldysh sigma-model
for time-dependent random Hamiltonians in Sec.~\ref{S:pert}.

At the {\em non-perturbative} level, the DM transformation (\ref{Q-DM})
can be considered as an alternative
parameterization of the complex manifold $Q^2=1$.
Therefore, changing variables from $Q$ to $b$ and $\bar b$ may be considered
as a contour deformation on the NLSM manifold which leaves the functional
integral invariant. We do not give a rigorous proof of this statement but
demonstrate the non-perturbative exactness of the DM transformation
on the example of the supersymmetric NLSM for unitary random matrices
in Sec.~\ref{S:GUE}.

Another advantage of the DM representation is that it provides
a compact algebraic way for deriving non-perturbative transfer-matrix
equations for the one-dimensional NLSM. We demonstrate this
with the diffusive supersymmetric sigma-model for quasi-one-dimensional
localization \cite{Efetov1983} where the use of the DM transformation
allows to derive the transfer-matrix Hamiltonian ``without fermions'',
using just the algebraic properties of the corresponding symmetric space,
see Sec.~\ref{S:Q1D}.

Finally, in Sec.~\ref{S:classes} we show that the DM transformation
introduced for the unitary symmetry class can be extended to other
unitary-like symmetry classes C and B/D introduced
by Altland and Zirnbauer\cite{Altland-Zirnbauer}
in the context of superconductivity.

The main facts about the Dyson-Maleev transformation
are summarized in Sec.~\ref{S:Summary}.

\section{Perturbation theory:
Keldysh sigma-model for time-dependent random Hamiltonians}
\label{S:pert}

We illustrate the utility of the DM formalism in the perturbative
regime with the problem of a quantum particle subject to
a time-dependent random unitary Hamiltonian $H(t)$. This problem
has been studied in the framework of the Keldysh sigma-model \cite{IS2006},
and the use of the DM parameterization has been shown to
greatly reduce the number of diagrams, in
comparison with the usual infinite-series
parameterization (\ref{Q-W}).

\subsection{The model}

In a semiclassical approximation, the dynamics of a particle subject
to quantum evolution with a non-stationary Hamiltonian $H(t)$
can be described as a diffusion process in the energy space.
Diffusive spreading of the wave function
is characterized by the diffusion coefficient $D$ which determines
the rate of the energy drift with time $t$:
\be
\label{en-dif}
  \overline{[E(t)-E(0)]^2}
  =
  2 \mls^3 D t ,
\ee
with $\mls$ being the mean level spacing.
The dimensionless diffusion coefficient $D$ depends on the rate
of variation of the Hamiltonian measured by the dimensionless
velocity of adiabatic energy levels:
\be
\label{alpha}
  \alpha
  =
  \frac{\pi}{\mls^4}
  \overline{\left(\frac{\partial E_n}{\partial t}\right)^2} .
\ee
In the limit $\alpha\gg1$, the energy absorption is due to transitions
in the continuous spectrum, and this regime can be
described by the linear-response Kubo
formula \cite{wilkinson,SimonsAltshuler93}.
In the opposite limit, $\alpha\ll1$, energy is absorbed during rare
Landau--Zener transitions between neighboring levels.
In these limiting cases, the diffusion coefficient $D(\alpha)$
has been obtained by Wilkinson \cite{wilkinson}:
\be
\label{D-asymp}
  D(\alpha)
  =
  \begin{cases}
    (\beta/2) \alpha, & \alpha\gg1 , \\
    c_\beta \alpha^{(\beta+2)/4} , & \alpha\ll1 ,
  \end{cases}
\ee
where $\beta=1,2$, and 4 for the orthogonal, unitary, and symplectic ensembles,
respectively, and $c_\beta$ are numerical coefficients.
Quite surprisingly, for the unitary ensemble, $c_2=1$
indicating that $D(\alpha)=\alpha$ both in the limits
of small and large $\alpha$.

Unfortunately, the methods used in Ref.~\onlinecite{wilkinson} to get
the asymptotics (\ref{D-asymp}) cannot be generalized to finite
values of $\alpha$. A general approach to calculating
the full dependence $D(\alpha)$ based on the Keldysh sigma-model
formalism has been derived in Ref.~\onlinecite{skvor}.
The field variable is the operator $Q$ which is the integral kernel
in time domain with values in $2\times 2$ matrices in
the Keldysh space (analogous to the retarded-advanced space in the
supersymmetric formalism).
The $Q$-matrix is subject to the constraint $Q^2=1$,
where a convolution over time arguments is implied.
As usual, the integration manifold is the orbit (\ref{orbit})
of $\Lambda_{tt'}=\delta_{tt'}\sigma_3$ (Pauli matrix in the Keldysh space)
under unitary rotations $U$.
In the dimensionless form (time measured in units of $\Delta^{-1}$),
the Keldysh action for the linearly driven random unitary Hamiltonian
takes the form \cite{skvor,IS2006}
\be
  S[Q]
  =
  - \frac{\pi}{2} \TrK
  \int dt\, (\partial_1 - \partial_2) Q_{t_1 t_2} \Big|_{t_1=t_2=t}
  +
  \frac{\pi\alpha}{4}
    \TrK \int\!\!\int dt\, dt'\, (t-t')^2\, Q_{tt'} Q_{t't} ,
\label{action}
\ee
where the coupling parameter $\alpha$ is defined in Eq.~(\ref{alpha}),
and $\TrK$ denotes the trace over the two-dimensional
Keldysh space.
The parameter $\alpha$ plays the role of the dimensionless conductance
which controls the strength of fluctuations around $\Lambda$:
they are small for $\alpha\gg1$ and strong for $\alpha\ll1$.


The NLSM (\ref{action}) contains full information
about the function $D(\alpha)$ which can be expressed as \cite{IS2006}
\be
  D(\alpha)
  =
  - \lim_{t\to\infty} \frac{1}{2t}
  \left. \frac{\partial^2}{\partial \eta^2} \right|_{\eta=0}
  {\mathcal D}_\eta(t) ,
\label{diffusion-coefficient}
\ee
where the diffuson ${\mathcal D}_{\eta}(t)$ is defined
through the correlation function of the off-diagonal components
$Q^{(\pm)}_{t_1 t_2}=\TrK (\sigma^\mp Q_{t_1 t_2})$
of the $Q$ field:
\be
  \corr{Q^{(+)}_{t_1,t_2} \, Q^{(-)}_{t_3,t_4}}
  =
  \int[DQ]\, e^{-S[Q]} Q^{(+)}_{t_1,t_2} \, Q^{(-)}_{t_3,t_4}
  =
  \frac{2}{\pi}
  \delta(t_1-t_2+t_3-t_4)
  {\mathcal D}_{t_1-t_2}(t_1-t_4) .
\label{full-diffuson-definition}
\ee

\begin{figure}
\epsfxsize=0.5\hsize
\centerline{\epsfbox{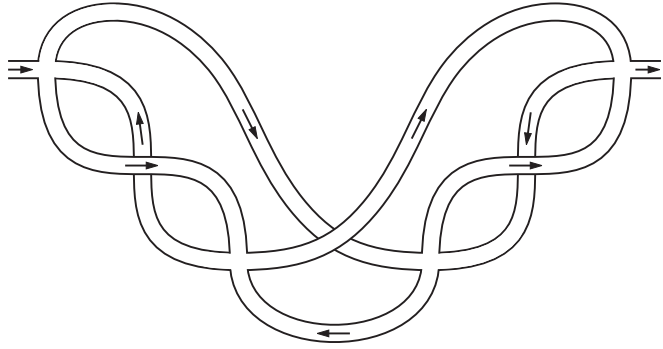}}
\caption{An example of the 4-loop diagram for the diffuson self-energy
(from Ref. \onlinecite{IS2006}).}
\label{fig:loop4example}
\end{figure}

\subsection{Dyson-Maleev transformation}

In the DM parameterization (\ref{Q-DM}), the Keldysh sigma-model
action (\ref{action}) takes the form
\be
  S[b,\bar{b}]
  =
  \frac{\pi}{2} \int\!\!\int dt_1\, dt_2\,
  \bar{b}_{12}
    \Big[ (\partial_1 + \partial_2) + \alpha(t_1-t_2)^2 \Big]
  b_{21}
  -
  \frac{\pi\alpha}{8} \int dt_1\, dt_2\, dt_3\, dt_4 \,
  (t_1-t_2)(t_3-t_4) \,
  b_{12}\bar{b}_{23}b_{34}\bar{b}_{41} .
\label{S-b}
\ee

The propagator of the quadratic part of the action (\ref{S-b}) is
given by
\be
  \corr{b_{t+\eta/2,t-\eta/2} \bar{b}_{t'-\eta'/2,t'+\eta'/2}}^{(0)}
  = \frac{2}{\pi}
  \delta(\eta-\eta')
  {\mathcal D}^{(0)}_{\eta}(t-t') ,
\label{bare-propagator}
\ee
where ${\mathcal D}^{(0)}_\eta(t)$ is the bare diffuson:
\be
  {\mathcal D}^{(0)}_\eta(t)
  =
  \theta(t) \exp[-\alpha\eta^2 t] .
\label{bare-diffuson}
\ee
Substituting (\ref{bare-diffuson}) into (\ref{diffusion-coefficient}),
one gets the large-$\alpha$ Kubo result $D(\alpha)=\alpha$.
Treating the quartic term in the action (\ref{S-b}) as a perturbation,
we can compute the perturbative series for
the diffuson ${\mathcal D}_{\eta}(t)$ and, consequently,
for $D(\alpha)$ in the limit of large $\alpha$.
From simple power counting \cite{skvor}, $L$-loop diagrams
give a contribution to $D(\alpha)$ proportional to $\alpha^{1-L/3}$.
For the unitary ensemble, the number of loops must be even,
which corresponds to expanding in powers of $\alpha^{-2/3}$:
\be
  D(\alpha) = \alpha
  \left(
    1 + \frac{d_2}{\pi^2\alpha^{2/3}} + \frac{d_4}{\pi^4\alpha^{4/3}} + \dots
  \right) .
\label{D-series}
\ee

In two loops, it has been found that $d_2=0$
(the nullification of this diagram is nontrivial, and
we are not aware of any symmetry reasons for this fact)\cite{skvor,SBK04}.
In the DM parameterization, there is just one two-loop diagram,
while the standard infinite-series parameterization (\ref{Q-W})
generates two distinct diagrams, one made of a 6-order vertex,
and another made of two 4-order vertices. While the DM parameterization
appears to be the most compact way of getting the analytic expression
in two loops, its use at this order of perturbation theory
is not crucial.

The coefficient $d_4$ is determined by four-loop diagrams.
Its calculation performed in Ref.~\onlinecite{IS2006} was only
possible with the use of the DM parameterization.
Indeed, four-loop diagrams in an arbitrary infinite-series
parameterization (\ref{Q-W}) would contain interaction vertices
of the $b$ and $\bar b$ fields up to the tenth order.
Then even a classification of the four-loop diagrams becomes
a sophisticated problem.
On the other hand, in the DM formalism with the
single quartic term in the action (\ref{S-b}),
there are only 20 irreducible four-loop diagrams for the diffuson
self-energy, see an example in Fig.~\ref{fig:loop4example}.
Some of them are related by symmetries,
and the total number of nonequivalent diagrams is 9.
They are given by seven-fold time integrals of the form
$$
  \int_0^\infty \!\!\!\!\dots \int_0^\infty
  \left( \prod_{i=1}^{7} dT_i \right)
  P_6 (T_1, \dots, T_7)
  e^{-S_3(T_1, \dots, T_7)} ,
$$
where $P_6(T_1, \dots, T_7)$
and $S_3(T_1, \dots, T_7)$ are
homogeneous polynomials of degrees six and three, respectively.
Numerical evaluation of these integrals shows that $d_4$
is zero within the accuracy of the calculation
and supplies the upper bound for its absolute
value: $|d_4|<3\times10^{-4}$.
A further discussion of this result
can be found in Ref.~\onlinecite{IS2006}.

To conclude this Section, we mention that the same diagram classification
applies to a wide class of NLSM with the action (\ref{S[Q]-general}),
whose Dyson-Maleev representation (\ref{S[Q]-DM}) contains only
the Gaussian and quartic terms. In particular, quantum dynamics
with an arbitrary time dependence of the Hamiltonian $H(t)$
can be described by the Keldysh action (\ref{action})
with the slightly modified last term\cite{skvor,BSK03},
thus also admitting the DM parameterization with
the quartic interaction \cite{IS2006}.
Finally, various versions of diffusive sigma-models also belong
to this class and have the same diagram classification in the DM
representation.

\section{Non-perturbative approach:
supersymmetric NLSM for unitary random matrices}
\label{S:GUE}

We start discussing the non-perturbative aspects of the DM transformation
with the simplest example of the supersymmetric NLSM for
the spectral statistics of unitary random matrices.

The theory is formulated in terms of the $4\times4$ supermatrix $Q$
acting in the direct product of the superspace (BF)
and the retarded-advanced space (RA).
The pair correlation function of the density of states,
$R(\omega)=\mls^2\corr{\rho(E)\rho(E+\omega)}$,
is given by the integral \cite{Efetov1983,Efetov-book}
\be
\label{R}
  R(\omega)
  =
  \frac1{16} \Re
  \int (\str k\Lambda Q)^2 e^{-S[Q]} DQ .
\ee
Here $\mls$ is the mean level spacing,
$\Lambda=\diag(1,-1)_{RA}$, and
$k=\diag(1,-1)_{BF}$ is the supersymmetry-breaking matrix.
The supertrace is defined as the bosonic trace minus fermionic trace.
The sigma-model action is
\be
\label{S[Q]-RMT}
  S[Q] = - \frac{i\pi\omega}{2\mls} \str \Lambda Q.
\ee

With the non-compact BB sector and compact FF sector,
the action (\ref{S[Q]-RMT}) possesses two saddle points:
the standard saddle point $Q=\Lambda$ (``north pole of the fermionic sphere'')
and $Q=\Lambda k$ (``south pole of the fermionic sphere''),
both belonging to the same connected component. The ``south pole''
is known to be responsible
for oscillations in the pair correlation function \cite{AA1995}.

The DM transformation (\ref{Q-DM}) may be considered as a special
change of variables which parameterizes almost the whole complex
manifold $Q^2=1$. However the ``south pole'' $\Lambda k$
does not correspond to any finite $b$ and $\bar b$ and can be achieved
only as a limiting point for $b=\diag(0,4/p)_{BF}$, $\bar b=\diag(0,p)_{BF}$
at $p\to0$.

In Efetov's parameterization \cite{Efetov1983,Efetov-book},
the integration contour over $Q$ is chosen in such a way
that the eigenvalues $\lambda_B$ and $\lambda_F$
of the block $Q^{RR}$ are real and bounded to the strip
\be
\label{strip}
  \lambda_B \geq 1 ,
  \qquad
  -1 \leq \lambda_F \leq 1
\ee
(the eigenvalues of the block $Q^{AA}$
are the opposite: $-\lambda_B$ and $-\lambda_F$).
Now a transition from Efetov's parameterization of the $Q$-manifold
to the DM parameterization can be realized as a deformation
of the integration contour in a multi-dimensional space.
It will be convenient to
impose an additional requirement that the eigenvalues of the block $Q^{RR}$
(and the block $Q^{AA}$ as well) do not change during the deformation.
Since the eigenvalues of the product $b\bar b$ are given by
$2(1-\lambda_B)$ and $2(1-\lambda_F)$, this requirement
constrains the eigenvalues of $b\bar b$ to be real and to belong
to the corresponding strip.
The constraint on the eigenvalues of $b\bar b$ is reminiscent of the truncation of the
bosonic Hilbert space in the original DM transformation for spin operators
(where it is formulated as the constraint on the eigenvalues of the
number of bosons $\hat a^\dagger \hat a$). We may also note a similar restriction
of the integration region in the theory of the DM
parameterization for classical ferromagnets \cite{Kolokolov2000}.

Throughout the deformation from Efetov's to DM parameterization,
we may maintain the unchanged relation between the commuting components of
$b$ and $\bbar$:
\be
  b
  =
  \begin{pmatrix}
    b_{11} & b_{12} \\
    b_{21} & b_{22}
  \end{pmatrix}_{BF}
  ,
\qquad
  \bar b
  =
  \begin{pmatrix}
    \bar b_{11} & \bar b_{12} \\
    \bar b_{21} & \bar b_{22}
  \end{pmatrix}_{BF}
  =
  \begin{pmatrix}
    -b_{11}^* & \bar b_{12} \\
    \bar b_{21} & b_{22}^*
  \end{pmatrix} .
\label{bb-elements}
\ee
This choice of the real section of the complex manifold $Q^2=1$ respects the
correct domains for $\lambda_B$ and $\lambda_F$.
The independent elements are the two complex numbers $b_{11}$ and $b_{22}$,
and the four Grassmann numbers $b_{12}$, $b_{21}$, $\bar b_{12}$, $\bar b_{21}$
(we do not use complex conjugation of Grassmann variables in our formalism).
An arbitrary integral over $Q$ can then be rewritten in the DM variables as
\be
\label{int-Q-DM}
  \int {\cal F}[Q] \, DQ
  =
  \int {\cal F}[Q(b,\bar b)] \,
  \theta(\lambda_B-1) \theta(1-\lambda_F) \theta(1+\lambda_F)
  \, Db \, D\bar b ,
\ee
where $\theta(x)$ is the step function,
and $Db \, D\bar b$ is the standard flat measure:
\be
  Db \, D\bar b
  =
  \frac{d\Re b_{11} \, d\Im b_{11} \, d\Re b_{22} \, d\Im b_{22}}{\pi^2}
  \,
  db_{12} \, db_{21} \, d\bar b_{12} \, d\bar b_{21}.
\label{bb-integration}
\ee
The integral (\ref{int-Q-DM}) may be calculated straightforwardly
in the variables $b_{ij}$ using the explicit expression for
the eigenvalues of the product $b\bar b$:
\begin{gather}
  2(1-\lambda_B)
  =
    b_{11} \bar b_{11}
  + b_{12} \bar b_{21}
  + \frac
    {(b_{11} \bar b_{12} + b_{12} \bar b_{22})(b_{21} \bar b_{11} + b_{22} \bar b_{21})}
    {b_{11} \bar b_{11} - b_{22} \bar b_{22} + b_{12} \bar b_{21} - b_{21} \bar b_{12}} ,
\\
  2(1-\lambda_F)
  =
    b_{22} \bar b_{22}
  + b_{21} \bar b_{12}
  + \frac
    {(b_{11} \bar b_{12} + b_{12} \bar b_{22})(b_{21} \bar b_{11} + b_{22} \bar b_{21})}
    {b_{11} \bar b_{11} - b_{22} \bar b_{22} + b_{12} \bar b_{21} - b_{21} \bar b_{12}} .
\end{gather}

These formulas allow a direct calculation of correlation functions in the DM
parameterization.

\subsection{Technicalities of the integration over $b$ and $\bar b$}

We illustrate the above approach by calculating the integral
\be
  Z
  =
  \int p(\str k \Lambda Q) F(\str \Lambda Q) \, DQ
  =
  \int p(\str k \Lambda Q) F(\str \Lambda Q) \,
  \theta(\lambda_B-1) \theta(1-\lambda_F) \theta(1+\lambda_F)
  Db \, D\bar b ,
\label{Z-def}
\ee
which for a suitable choice of the functions $p$ and $F$ gives the
level-level correlation function (\ref{R}).

Note that in the flat parameterization used,
$\str k \Lambda Q = 4 - (1/2) \str k(b\bar b+\bar bb) =
4 - b_{11}\bar b_{11} - b_{22}\bar b_{22}$
does not depend on Grassmann variables.
A direct calculation (expanding in Grassmann variables)
shows, quite surprisingly, that the same is true for two step
functions in (\ref{Z-def}):
$\theta(\lambda_B-1)=\theta(-b_{11}\bar b_{11})$,
$\theta(1-\lambda_F)=\theta(b_{22}\bar b_{22})$.
Therefore, only two terms in the integrand (\ref{Z-def}) contain
Grassmann variables: $F(\str \Lambda Q)=F(\str b\bar b)$
and $\theta(1+\lambda_F)$.
Expanding these functions in Grassmann variables and evaluating
the integral over them, after a straightforward but lengthy
calculation we arrive at
\be
\label{Z}
  Z
  =
  4 \int_1^\infty d\lambda_B \int_{-1}^1 d\lambda_F \:
  p''(2\lambda_B+2\lambda_F) F(2\lambda_B-2\lambda_F)
  +
  p(4) F(0) ,
\ee
where the second term is the anomaly contribution.
In particular, it is responsible for the proper normalization
$Z=F(0)$ for $p=1$.

To find the level-level correlator (\ref{R}), we should
use (\ref{Z}) with $p(x)=x^2/16$ and $F(x)=\exp(i\pi\omega x/2\mls)$.
Evaluating the integrals over $\lambda_B$ and $\lambda_F$ we arrive
at the well-known result \cite{Mehta}
\be
\label{R-GUE}
  R(\omega) = 1 - \frac{\sin^2(\pi\omega/\mls)}{(\pi\omega/\mls)^2} .
\ee

By reproducing the standard expression (\ref{R-GUE}) we demonstrate
that restricting the integration region over $b$ and $\bar b$
to the strip (\ref{strip}) makes the DM transformation
{\em non-perturbatively}\/ equivalent to the initial NLSM
in the $Q$-representation.

To be precise, our example (\ref{Z-def}) involve only the diagonal
blocks $Q^{RR}$ and $Q^{AA}$. Therefore, our derivation is equivalent to that
in the Hermitian ``square-root-odd'' parameterization \cite{YurLerner99,IS2006}
of the type (\ref{Q-W}) with $f(W) = 1+W^2/2 + W \sqrt{1+W^2/4}$
which has unit Jacobian. This parameterization has the same
diagonal blocks as the DM one and the same integration measure.

In fact, a continuous interpolation is possible between
the ``square-root-odd'' and DM parameterization, such that the
diagonal blocks $Q^{RR}$ and $Q^{AA}$ remain unchanged, and the
Jacobian equals one throughout the interpolation\cite{Sekatski}.

The power of the DM parameterization will come in the full glory
in the next Section when the off-diagonal blocks of the $Q$
matrix will play the role.

\section{Non-perturbative approach:
supersymmetric NLSM for quasi-one-dimensional localization}
\label{S:Q1D}

In this section, we apply the Dyson--Maleev parameterization
to the transfer-matrix treatment of the Efetov supersymmetric
sigma-model for quasi-one-dimensional disordered systems.
While most of the results  of this section reproduce the known
ones\cite{EL83,Efetov-book}, we find it instructive to
rederive the transfer-matrix formulas with the DM method for two
reasons:
\begin{list}{$\bullet$}{}
\item
the derivation of the transfer-matrix Hamiltonian simplifies in the
DM parameterization and does not involve Grassmann variables, but uses
algebraic relations for supermatrices instead;
\item
the derivation of expressions for the correlation functions uses
more transparently the symmetries of the sigma-model, and the
intermediate states are more easily classified by the appropriate
representations of the symmetry supergroup.
\end{list}

The main goal of the calculation in this section is obtaining a closed
expression for the two-point correlation function in the one-dimensional
sigma-model~\cite{Efetov-book}:
\begin{equation}
\int [{\cal D}Q]\;  \str(A_1 Q(x_1)) \;  \str(A_2 Q(x_2))\;
e^{-S[Q]}\, ,
\label{sigma-model-path-integral}
\end{equation}
where $A_1$ and $A_2$ are arbitrary supermatrices. The sigma-model
action is
\be
  S[Q] = - \frac{\pi\nu_1}{4} \str \int dx \,
\left[D \left(\frac{dQ}{dx}\right)^2
+ 2i\omega \Lambda Q \right] ,
\label{sigma-model-action}
\ee
where $\nu_1=\nu S_W$ is the q1D density of states ($\nu$ is the 3D density
of states and $S_W$ is the area of the wire), $D$ is the diffusion
coefficient, $\omega$ is the energy difference between the retarded
and advanced sectors. The supertrace $\str()$ is
defined in the convention of Refs.~\onlinecite{VWZ,Mirlin2000}: bosonic trace
minus fermionic trace. We consider the unitary symmetry class, and $Q$ is the
supermatrix of dimension $2|2$: it has the $2\times 2$ structure in the
retarded-advanced space. It is subject to the sigma-model constraint
$Q^2=1$ and is obtained by rotating the ``north pole'' (\ref{Lambda-definition})
[independently at every position $x$].

First, we rederive the mapping of the path integral
(\ref{sigma-model-path-integral}) onto a finite-dimensional
quantum mechanics (the transfer-matrix approach)\cite{EL83},
and then we express it
in terms of dynamic correlation functions
in this quantum mechanics.

\subsection{Transfer-matrix Hamiltonian. General formalism.}

It will be convenient to rescale the coordinate $x$ in
(\ref{sigma-model-action}) by the localization length
$\xi=2\pi\nu_1 D$.
The dimensionless length $\tau=x/\xi$ will play the role of
(imaginary) time in the quantum mechanics.
In the rescaled variables, the action is
\be
  S
  = \str \int d\tau \,
  \left[ -\frac{\dot{Q}^2}{8} + \Omega \Lambda Q \right] ,
\ee
where $\Omega = -i\omega\xi^2/(4D)$.

With the use of the DM parameterization (\ref{Q-DM}), the action may be
rewritten as:
\be
S=\int d\tau \, L\, , \qquad
L= - \frac{1}{4} \str \left[\dot b \dot\bbar +
\frac{1}{4}\dot{b} \bbar \dot{b} \bbar  + 4 \Omega b\bbar \right] \, .
\ee
This action can be converted into the transfer matrix (Hamiltonian)
with the usual Legendre transformation. The
momentum variables conjugate to $b$ and $\bbar$ are
\be
\Pi = \frac{\partial L}{\partial \dot{b}} =
\frac{1}{4}\left[ \dot\bbar + \frac{1}{2} \bbar \dot{b} \bbar \right] \, ,
\qquad
\bar\Pi =\frac{\partial L}{\partial{\dot\bbar}} =
\frac{1}{4} \dot{b}
\ee
(here $b$, $\bbar$, $\Pi$, and $\bar\Pi$ are supermatrices, and their
products involve the usual index convolutions), and
the transfer matrix $H$ is given by the Legendre transform
\be
H=L - \str\left(\Pi \dot b + \bar\Pi \dot\bbar \right) =
\Delta_{-1} + \Delta_0 + U\, ,
\label{transfer-matrix-1}
\ee
where
\be
\Delta_{-1}=
4 \str\left( \Pi \bar\Pi\right)\, , \qquad
\Delta_0 =
 - \str\left( \bar\Pi \bbar \bar\Pi \bbar \right)\, , \qquad
U=
- \Omega \str (b \bbar)\, .
\label{transfer-matrix-2}
\ee
($\Delta_0$ preserves the degree of $b$ and $\bbar$,
$\hat\Delta_{-1}$ lowers the degrees by one, and $U$ raises the degrees
by one).

This Hamiltonian should now be understood as a quantum Hamiltonian
with the canonical expressions for the momentum variables:
\be
\Pi=\frac{\partial}{\partial b}, \qquad
\bar\Pi =\frac{\partial}{\partial\bbar} .
\ee
Since $\bbar$ and $\bar\Pi$ do not commute, a care is needed for
a proper ordering of the quartic term $\Delta_0$ of the Hamiltonian.
However, in our example of the supersymmetric sigma-model,
the commutators of $\bbar$ and $\bar\Pi$ in the matrix product
in (\ref{transfer-matrix-2}) generate terms proportional to
$\str \bm{1} =0 $, therefore ordering is not important. The
same situation occurs in the replica and Keldysh sigma-models,
which are also defined so that $\tr \bm{1}=0$.  A more
careful consideration may, however, be necessary if one wishes
to extend this construction to other sigma-models (with non-normalized
partition functions).

The above formal derivation is applicable to any sigma-model
with the Dyson--Maleev parameterization. We further specify the
explicit form of the Hamiltonian by
\begin{list}{$\bullet$}{}
\item
projecting the Hamiltonian onto a particular representation of
the symmetry group of the sigma-model action (\ref{sigma-model-action});
\item
restricting $b$ and $\bbar$ to belong to a particular finite-dimensional
space.
\end{list}

\subsection{Hamiltonian in the invariant (``singlet'') symmetry sector}

The sigma-model action (and hence the Hamiltonian $H$) is invariant
with respect to the supergroup $H_R \times H_A$ of
independent rotations in advanced and retarded sectors, as
described in Appendix \ref{Appendix:math}. In this subsection, we
consider the restriction of the Hamiltonian
(\ref{transfer-matrix-1})--(\ref{transfer-matrix-2}) onto the sector
of states invariant under the $H_R \times H_A$ group (the ``singlet''
sector).

In this paper we keep the symmetry analysis at a simple level
and refer the interested reader to Ref.~\onlinecite{supergroup-rep},
which contains a helpful introduction into the representation theory
of the supergroup $GL(1|1)$ relevant for our example.

The ``singlet'' wave functions may generally be written as functions
of cyclic traces $x_n=\str (b \bbar)^n$. When projected onto
such singlet functions, the Hamiltonian takes the form
\begin{gather}
  \Delta_0 \to \Delta_0^{(S)} = -\sum_{k_1,k_2=1}^{\infty}
  \left[
  k_1 k_2 \;  x_{k_1+k_2}\frac{\partial}{\partial x_{k_1}}
  \frac{\partial}{\partial x_{k_2}}
  +
  (k_1+k_2) \; x_{k_1} x_{k_2}
  \frac{\partial}{\partial x_{k_1+k_2}}
  \right] ,
\label{singlet-Delta-0}
\\
  \Delta_{-1} \to \Delta_{-1}^{(S)} = 4 \sum_{k_1,k_2=1}^{\infty}
  \left[
  k_1 k_2 \; x_{k_1+k_2-1}
  \frac{\partial}{\partial x_{k_1}} \frac{\partial}{\partial x_{k_2}}
  +
  (k_1+k_2+1) \; x_{k_1} x_{k_2}
  \frac{\partial}{\partial x_{k_1+k_2+1}}
  \right] ,
\label{singlet-Delta-minus}
\\
  U \to U^{(S)} = -\Omega x_1 .
\label{singlet-U}
\end{gather}

Note that so far the formalism applies generally to
arbitrary sigma-models admitting the DM parameterization.

\subsection{Reducing the Hamiltonian to a finite-dimensional space}

Now we make the second step of specifying the particular type
of the finite-dimensional symmetric (super)space. This will
be done with the so-called ``closing relations''.

In our example, with the matrix $Q$ having the dimension $2|2$,
the matrices $b$ and $\bbar$ have the dimension $1|1$, and so does
their product $b \bbar$. For this finite-dimensional space of matrices,
among the infinite set of variables $x_i$, only the first two
are independent. All the higher variables $x_i$ may be expressed in terms
of $x_1$ and $x_2$ with the help of ``closing relations'' which carry
information about the properties of the particular
symmetric superspace.
When the Hamiltonian (\ref{transfer-matrix-1}),
(\ref{singlet-Delta-0})--(\ref{singlet-U}) acts on a wave function
of the form $\Psi(x_1,x_2)$, only two higher-order variables are
generated: $x_3$ and $x_4$. Thus it is sufficient to use the two
closing relations for the space of $(1|1)$ supermatrices:
\be
x_3=\frac{1}{4x_1}\left(3 x_2^2 + x_1^4 \right)\, , \qquad
x_4=\frac{1}{2}\left(x_1^2 x_2 + \frac{x_2^3}{x_1^2}\right) .
\label{singlet-closing}
\ee

For future discussion of $RA$ sector, it is helpful to note
that those two relations are consequences of
one matrix relation (the characteristic polynomial):
\be
(b \bbar)^2 - \left(\frac{x_2}{x_1}\right) b\bbar +
\frac{1}{4}\left[\left(\frac{x_2}{x_1}\right)^2 - x_1^2 \right] = 0 .
\label{closing-matrix}
\ee

Using the closing relations (\ref{singlet-closing}), we reduce the
operators (\ref{singlet-Delta-0}) and (\ref{singlet-Delta-minus}) to
the differential operators acting on the functions of two variables
$\Psi(x_1,x_2)$:
\begin{gather}
\Delta_0^{(S)} \to
\Delta_0^{(S,1|1)} = - \left[x_2 \left(\frac{\partial}{\partial x_1}\right)^2
+ \left(\frac{3x_2^2}{x_1}+x_1^3\right) \frac{\partial}{\partial x_1}
\frac{\partial}{\partial x_2} + 2 \left(x_1^2 x_2 + \frac{x_2^3}{x_1^2}\right)
\left(\frac{\partial}{\partial x_2}\right)^2 +
2 x_1^2 \frac{\partial}{\partial x_2} \right]\, ,
\label{singlet-11-Delta-zero}
\\
\Delta_{-1}^{(S)} \to
\Delta_{-1}^{(S,1|1)}= 4\left[
x_1 \left(\frac{\partial}{\partial x_1}\right)^2 +
4 x_2 \frac{\partial}{\partial x_1} \frac{\partial}{\partial x_2} +
\left(\frac{3 x_2^2}{x_1} + x_1^3 \right)
\left(\frac{\partial}{\partial x_2}\right)^2
\right]\, .
\label{singlet-11-Delta-minus}
\end{gather}
Finally, it is convenient to change the variables to
the eigenvalues $\lambda_B$ and $\lambda_F$ of the matrix $Q^{RR}$
(or, equivalently, of the matrix $-Q^{AA}$).
The matrix $b\bbar$ then has the eigenvalues $2(1-\lambda_B)$ and
$2(1-\lambda_F)$, which produces the change-of-variable relations
\be
x_1=2(\lambda_F-\lambda_B)\, , \qquad
x_2=4(\lambda_F-\lambda_B)(2-\lambda_F-\lambda_B) .
\label{x-to-lambda}
\ee
With this change of variables, the Hamiltonian
\be
H^{(S,1|1)} = \Delta_0^{(S,1|1)} +  \Delta_{-1}^{(S,1|1)} + U^{(S)}
\ee
is equivalent to that in Efetov's Eq.~(11.38) in Ref.~\onlinecite{Efetov-book}.

\subsection{Bilinear forms and self-conjugate singlet Hamiltonian}

The Hamiltonian derived above is not self-conjugate with respect to the
flat integration in $\lambda_B$ and $\lambda_F$. For the convenience
of interpreting it as a quantum-mechanical Hamiltonian, we will use
an appropriate Jacobian to make it self-conjugate.

For this step of the derivation, as well as for the future use, we will
recall some well-known properties of the supersymmetric formalism.

The ``natural'' bilinear form on wavefunctions $\Psi(Q)$ can be defined as
\be
\langle \Psi_1(Q) | \Psi_2(Q) \rangle = \int DQ\, \Psi_1(Q)\, \Psi_2(Q)
=\int Db\, D\bbar\, \Psi_1 \Psi_2 ,
\label{Q-form}
\ee
where the integration is performed over the finite-dimensional superspace
of $Q$ with the normalization $\int DQ = 1$. The second equality reflects
the fact that the Jacobian of the DM parameterization equals one. The
integration measure over $Db\, D\bbar$ is defined in (\ref{bb-elements}),
(\ref{bb-integration}).

When restricted to the ``singlet'' states, this bilinear form is degenerate:
it is given by the ``anomaly''
\cite{Efetov-book}
\be
\langle \Psi_1(\lambda_B,\lambda_F) | \Psi_2(\lambda_B,\lambda_F) \rangle
= \Psi_1 \Psi_2 \Big|_{\lambda_B=\lambda_F=1} .
\ee
The anomaly gives 1 for the ground state and 0 for the excited states.
To define a nondegenerate bilinear form on the excited states, we can
define the ``renormalized'' form
\be
\langle \Psi_1(\lambda_B,\lambda_F) | \Psi_2(\lambda_B,\lambda_F) \rangle_*
= \int d\lambda_F\; d\lambda_B\;
\Psi_1 \Psi_2\; J
\ee
with the Jacobian
\be
  J=\frac{1}{(\lambda_B-\lambda_F)^2} .
\label{Jacobian}
\ee
This bilinear form $\langle \cdot | \cdot \rangle_*$ has the following
properties:
\begin{list}{$\bullet$}{}
\item
with respect to this form, the Hamiltonian $H^{(S,1|1)}$
is self-conjugate. In fact, the Jacobian
(\ref{Jacobian}) can be obtained uniquely (up to an overall normalization)
from this condition;
\item
it is well-defined and non-degenerate on singlet {\it excited}\/ states;
\item
it is divergent on the ground state, and its divergence is proportional to
$\langle \cdot | \cdot \rangle$.
\end{list}

We will also use the ``flat''
bilinear form on singlet states:
\be
\llangle \Psi_1(\lambda_B,\lambda_F) | \Psi_2(\lambda_B,\lambda_F) \rrangle =
\int d\lambda_F\; d\lambda_B\; \Psi_1 \Psi_2.
\label{flat-form}
\ee

To produce a Hamiltonian self-conjugate with respect to this flat
scalar product, it is sufficient to redefine the wave function by
\be
f(\lambda_B,\lambda_F)=\sqrt{J}\; \Psi(\lambda_B,\lambda_F)\, , \qquad
\sqrt{J}=\frac{1}{\lambda_B-\lambda_F}
\label{f-function}
\ee
Remarkably, in terms of the new wave functions
$f(\lambda_B,\lambda_F)$, the Hamiltonian takes a very simple form,
with the bosonic and fermionic degrees of freedom separated:
\be
\tilde H_0 = \sqrt{J} H^{(S,1|1)} \frac{1}{\sqrt{J}} =
\tilde H (\lambda_B) - \tilde H (\lambda_F)
\label{Hamiltonian-singlet-1}
\ee
where
\be
\tilde H (\lambda) = \frac{\partial}{\partial\lambda}
(1-\lambda^2) \frac{\partial}{\partial\lambda} + 2 \Omega \lambda
\label{Hamiltonian-singlet-2}
\ee
(the same expression in the bosonic and fermionic sectors).
This expression is convenient to compare with
Eq.~(11.38) in Ref.~\onlinecite{Efetov-book}.

The separation of variables implies that the {\it excited}\/
eigenstates have the product form $f_B(\lambda_B) f_F(\lambda_F)$ with
the energies $E^{(B)} - E^{(F)}$,
where $E^{(B)}$ and $E^{(F)}$ are the eigenvalues of the Hamiltonians
$\tilde H (\lambda_B)$ and $\tilde H (\lambda_F)$, respectively
(even though those
two Hamiltonians have identical algebraic form, they are defined on different
intervals $\lambda_B\in [1;\infty)$ and $\lambda_F\in[-1;1]$, and therefore
have very different spectra).

The separation of variables does not apply, however, to the {\it ground}\/
state: it has the boundary condition $\Psi(\lambda_B=\lambda_F=1)=1$,
and the corresponding wave function
$f(\lambda_F,\lambda_B)$ given by (\ref{f-function}) is not normalizable.

\subsection{Separation of variables for the ground state}

It has been noted in Ref.~\onlinecite{Skvortsov-Ostrovsky} that
the ground state of the Hamiltonian (\ref{Hamiltonian-singlet-1})
can also be obtained with a separation of variables, but with
a {\it different}\/ one (only applicable to the ground state).
We would like therefore to make a short
deviation from the main line of our paper (DM parameterization)
to briefly review the construction of the ground state (in a
slightly different way from the original derivation in
Ref.~\onlinecite{Skvortsov-Ostrovsky}).

If one defines
\be
D_-=\frac{\partial}{\partial\lambda_B} - \frac{\partial}{\partial\lambda_F}
\, ,
\ee
then a remarkable algebraic identity follows:
\be
\tilde H_0 \sqrt{J} D_- = D_- \sqrt{J} \tilde H_0^{\#}
\ee
where the variables separate again
in the ``dressed'' Hamiltonian $\tilde H_0^{\#}$:
\be
\tilde H_0^{\#} =
\tilde H^{\#} (\lambda_B) - \tilde H^{\#} (\lambda_F)\, ,
\label{Hamiltonian-singlet-reordered-1}
\ee
with
\be
\tilde H^{\#} (\lambda) =
(1-\lambda^2) \left( \frac{\partial}{\partial\lambda} \right)^2
+ 2 \Omega \lambda .
\label{Hamiltonian-singlet-reordered-2}
\ee
With this algebraic trick, the zero-energy ground state of $\tilde H_0$
is obtained from the zero-energy state of $\tilde H^{\#}_0$ by applying the
``dressing'' operator $\sqrt{J} D_-$,
\be
f_0 = \sqrt{J} D_- \Psi^{(B)}_E(\lambda_B) \Psi^{(F)}_E(\lambda_F) ,
\label{zero-state-1}
\ee
where $\Psi^{(B)}_E(\lambda_B)$ and $\Psi^{(F)}_E(\lambda_F)$ are
the bosonic and fermionic eigenstates of $\tilde H^{\#} (\lambda_B)$ and
$\tilde H^{\#} (\lambda_F)$ at the same energy $E$. One can verify that
the spectrum of  $\tilde H^{\#} (\lambda_B)$ is bounded by $2\Omega$ from
below and that that of $\tilde H^{\#} (\lambda_F)$ is bounded by $2\Omega$ from
above, and therefore the only possibility to cancel the bosonic
and fermionic energies in (\ref{zero-state-1})
is to take the bosonic and fermionic eigenstates at the same energy
$E=2\Omega$. Those eigenstates are given by modified Bessel functions,
which results in the following explicit expression for the
zero mode\cite{Skvortsov-Ostrovsky} [with the proper normalization
$\Psi_0(\lambda_B=\lambda_F=1)=1$]:
\be
\Psi_0=\frac{1}{2\Omega \sqrt{J}} f_0 =
\frac{1}{2\Omega} D_- \left[ p_B K_1(p_B) p_F I_1(p_F)\right]\, ,
\label{zero-state-2}
\ee
where $p_B=2\sqrt{2\Omega(1+\lambda_B)}$, $p_F=2\sqrt{2\Omega(1+\lambda_F)}$.

This alternative derivation of the ground-state wave function suggests
the presence of a certain algebraic structure of the Hamiltonian related
to the supersymmetry. Unfortunately, at present we are unable to
properly identify this structure and leave this interesting question
for future studies.

\subsection{Matrix-element reduction rules}

Our calculations in the DM parameterization may be simplified with
the help of relations connecting the expectation values of
non-singlet combinations of $b$ and $\bbar$ to a singlet
bilinear form in the coordinates $\lambda_B$ and $\lambda_F$.
These relations are somewhat similar to the Wigner-Eckart theorem,
with the role of the symmetry group played by the supersymmetry
$H_R\times H_A$ of the action.

For the $Q^{RR}$--$Q^{AA}$ correlation functions (discussed in the
following subsection), we will need the following relation:
\begin{equation}
\langle\Psi_1 | \str(k_R b \bbar)\, \str(k_A \bbar b) | \Psi_2 \rangle
= 4\, \str k_R \, \str k_A\, \llangle \Psi_1 | \Psi_2 \rrangle
\label{magic-singlet}
\end{equation}
for any two $H_R\times H_A$-invariant states $\Psi_1(\lambda_B,\lambda_F)$ and
$\Psi_2(\lambda_B,\lambda_F)$ and for any two supermatrices $k_R$ and $k_A$.
In this relation, the left-hand side involves the bilinear form (\ref{Q-form}),
and the right-hand side --- the flat integration (\ref{flat-form}).

This relation may be established by an explicit calculation (this is the
only calculation in this section explicitly involving the Grassmann
variables). In a somewhat different form, this relation has been
used in Ref.~\onlinecite{Skvortsov-Ostrovsky} for calculating
local correlations $\langle Q^{RR}(0) Q^{AA}(0) \rangle$.

For the $Q^{RA}$--$Q^{AR}$ correlation functions (see discussion
in subsection \ref{SS:off-diagonal}),
we will need two other relations:
\begin{equation}
\langle\Psi_1 | \str(k_{RA}  \bbar)\, \str(k_{AR} b) | \Psi_2 \rangle
= \str (k_{RA} k_{AR}) \, \llangle \Psi_1 | M_1
| \Psi_2 \rrangle
\label{magic-triplet-1}
\end{equation}
and
\begin{equation}
\langle\Psi_1 | \str(k_{RA}  \bbar)\, \str(k_{AR} b \bbar b) | \Psi_2 \rangle
= \str (k_{RA} k_{AR}) \, \llangle \Psi_1 | M_2
| \Psi_2 \rrangle
\label{magic-triplet-2}
\end{equation}
for any two $H_R\times H_A$-invariant states $\Psi_1$ and
$\Psi_2$ and for any two supermatrices $k_{RA}$ and $k_{AR}$.
Here $M_1$ and $M_2$ are some functions of $\lambda_B$ and $\lambda_F$.
A direct calculation gives
\be
M_1=-\frac{2}{\lambda_B-\lambda_F}\, ,\qquad
M_2=4\frac{\lambda_B+\lambda_F-2}{\lambda_B-\lambda_F} .
\ee

\subsection{Diagonal two-point correlations}

We now turn to the calculation of the correlation functions
(\ref{sigma-model-path-integral}). From symmetry considerations, one
easily finds that the only nontrivial correlations are those
between $RR$--$AA$ and between $RA$--$AR$ blocks:
\be
\langle \str[k_R Q^{RR}(0)] \str[k_A Q^{AA}(\tau)] \rangle
\qquad \text{and} \qquad
\langle \str[k_{RA} Q^{RA}(0)] \str[k_{AR} Q^{AR}(\tau)] \rangle ,
\ee
where $k_R$, $k_A$, $k_{RA}$, and $k_{AR}$ are some arbitrary supermatrices.
The $RR$--$AA$ correlations will be further referred to as ``diagonal''
and $RA$--$AR$ as ``off-diagonal''.

At $\tau=0$, these correlations can be easily found from the explicit form
of the ground state $\Psi_0$, together with the
reduction rules discussed in the previous
subsection. The results are known from
Ref.~\onlinecite{Skvortsov-Ostrovsky}. At $\tau > 0$, one needs to insert
the appropriate evolution operator between the times $0$ and $\tau$.

We will now specifically discuss the case of diagonal correlations.
In this case, one of the two observables, $Q^{RR}$, is invariant with respect
to the $H_A$ rotations, while the other observable, $Q^{AA}$ is
invariant with respect to the $H_R$ rotations. Therefore one finds that
all the intermediate excited states generated between $0$ and $\tau$ should
be singlets, and one can reduce the problem to calculating a certain
expectation value of the singlet evolution operator
$\exp(-\tilde H_0 \tau)$.

A formal way to prove this statement is the following.
First, we rewrite the $RR$--$AA$ correlation function as
\be
\langle \str[k_R Q^{RR}(0)] \str[k_A Q^{AA}(\tau)] \rangle=
\langle \Psi_0 | \str[k_A Q^{AA}] \exp(-H\tau) \str[k_R Q^{RR}] |
\Psi_0 \rangle .
\ee
The state $Q^{RR} \Psi_0$ is a quadruplet of states of the form
\be
\Psi={\bf 1} \psi_0 + b\bbar \psi_1\, ,
\label{RR-general}
\ee
 where $\psi_n$ are some
$H_R\times H_A$-invariant states. Under evolution with the Hamiltonian
$H$ given by (\ref{transfer-matrix-1})--(\ref{transfer-matrix-2}),
this quadruplet evolves within a subspace
spanned by the states  $(b\bbar)^n \psi_n$. In our 1+1-dimensional example,
the higher powers of $b\bbar$ can be expressed via the first two with the
relation (\ref{closing-matrix}). Therefore, the quadruplet
$\exp(-H\tau) Q^{RR} \Psi_0$ can also be represented in the form
(\ref{RR-general}). In principle, with the technique of the previous
subsection, we can project the Hamiltonian onto such states and derive the
corresponding differential $2\times 2$ Hamiltonian acting on the functions
$(\psi_0,\psi_1)$. The projected Hamiltonian is obviously triangular:
it preserves the subspace of $H_R\times H_A$-invariant states ${\bf 1} \psi_0$.
In this subspace we can define a basis of eigenstates of $H$ consisting of the
ground state $\Psi_0$ (with energy zero) and the excited states
\begin{equation}
\Psi_{kk'}=\str(b\bbar) f_{kk'}\, ,
\qquad
f_{kk'}=\Psi_k^{(B)}(\lambda_B) \Psi_{k'}^{(F)}(\lambda_F)
\label{fkkprime}
\end{equation}
with the energies $E_{kk'}=E_k^{(B)}-E_{k'}^{(F)}$.
We can complement this basis to
the eigenstate basis in the space of states (\ref{RR-general}). Note that for
any eigenstate quadruplet of the form (\ref{RR-general}), its supertrace
should also be an eigenstate. Therefore, the eigenstates
$(\psi_0,\psi_1)$ have the form
$\psi_1=f_{kk'}$, $\psi_0=g_{kk'}$ with $f_{kk'}$ given by (\ref{fkkprime})
and with some functions $g_{kk'}$ whose
exact form is of no importance for our calculation. Now we can quite
generally decompose
\begin{equation}
Q^{RR} \Psi_0= \sum_{kk'} c_{kk'}[g_{kk'}+ b\bbar f_{kk'}]+
\sum_{kk'} d_{kk'} \str(b\bbar) f_{kk'} + e \Psi_0
\label{RR-decomposition}
\end{equation}
with some coefficients $c_{kk'}$, $d_{kk'}$, and $e$.
By taking a convolution of this expression with (properly normalized)
$\Psi_0$, we find $e=1$.
We further apply the evolution operator to this expression and take the
convolution with $Q^{AA} \Psi_0$. Only the terms containing $c_{kk'}^2$
and $e^2$ survive, and, using (\ref{magic-singlet}), we find
\begin{equation}
\langle \str[k_R Q^{RR}(0)] \str[k_A Q^{AA}(\tau)] \rangle=
- \str k_A \, \str k_R\, \left[ 4 \sum_{kk'} c_{kk'}^2 e^{-E_{kk'}\tau}
\llangle f_{kk'} | f_{kk'} \rrangle + 1 \right]\, .
\end{equation}
Finally, observing that the coefficients $c_{kk'}$
in (\ref{RR-decomposition}) may be obtained by decomposing $\Psi_0$ in
the basis $f_{kk'}$ of eigenstates of the Hamiltonian
(\ref{Hamiltonian-singlet-1})--(\ref{Hamiltonian-singlet-2}), we
arrive at the final result of this calculation
\begin{equation}
\langle \str[k_R Q^{RR}(0)] \str[k_A Q^{AA}(\tau)] \rangle
= - \str k_A \, \str k_R\,
 \left[ \llangle \Psi_0 | e^{-\tilde H_0 \tau} | \Psi_0 \rrangle
+ 1 \right] \, .
\label{singlet-evolution}
\end{equation}
At $\tau=0$, this expression agrees with the results of
Ref.~\onlinecite{Skvortsov-Ostrovsky}.

\subsection{Off-diagonal correlations}
\label{SS:off-diagonal}

In a similar way, one can derive an expression for the
off-diagonal two-point correlation function. The main difference is that
now the intermediate excited states are of the form
\begin{equation}
\Psi= b\phi_0 + b\bbar b \phi_1
\label{RA-general}
\end{equation}
(with $H_R\times H_A$-invariant $\phi_0$ and $\phi_1$), instead
of (\ref{RR-general}). Both components $\phi_0$ and $\phi_1$ are
coupled to each other by the Hamiltonian, and thus we
obtain an expression analogous to (\ref{singlet-evolution}), but
with the effective Hamiltonian having an additional 2$\times$2 matrix
structure (in the $\phi_0$--$\phi_1$ basis).

From the same matrix closing relation (\ref{closing-matrix}), we can
obtain the closing relations in the $RA$ sector:
\begin{eqnarray}
b(\bbar b)^2 &=& \left( \frac{x_2}{x_1} \right) b \bbar b
 + \frac{x_1^4 - x_2^2}{4 x_1^2} b \\
b(\bbar b)^3 &=& \frac{x_1^4 + 3 x_2^2}{4 x_1^2} b \bbar b
 +  \frac{x_2(x_1^4 - x_2^2)}{4 x_1^3} b \nonumber
\end{eqnarray}
By using these closing relations, the Hamiltonian (\ref{transfer-matrix-1})
projected onto the states (\ref{RA-general}) may be represented
as a 2$\times$2 matrix of differential operators acting on the vectors
$(\phi_0, \phi_1)$:
\begin{equation}
H^{(RA,1|1)}=
H^{(S,1|1)}+
\begin{pmatrix}
4 && \frac{x_2^2 - x_1^4}{2 x_1^2} \cr
0 && 12 - 2 \frac{x_2}{x_1}
\end{pmatrix}
\frac{\partial}{\partial x_1} +
\begin{pmatrix}
0 && \frac{(x_2^2 - x_1^4)x_2}{x_1^3} + 6 \frac{x_1^4 - x_2^2}{x_1^2} \cr
8 && 24 \frac{x_2}{x_1} - \frac{x_1^4 + 3 x_2^2}{x_1^2}
\end{pmatrix}
\frac{\partial}{\partial x_2} \, ,
\end{equation}
where $H^{(S,1|1)}$ is the Hamiltonian in the singlet sector
multiplied by the unit 2$\times$2 matrix.

As in the singlet sector, we rewrite this Hamiltonian in the variables
$\lambda_B$ and $\lambda_F$ [using (\ref{x-to-lambda})] and rotate
it to a self-conjugate form:
\begin{equation}
\tilde H_{RA} = U^{-1} H^{(RA,1|1)} U \, , \qquad
U=
\begin{pmatrix}
2(\lambda_B-1) && 2(\lambda_F-1) \cr
1 && 1
\end{pmatrix}
\begin{pmatrix}
\varphi_F && 0 \cr
0 && \varphi_B
\end{pmatrix} \, ,
\qquad
\varphi_F=\sqrt\frac{1+\lambda_F}{1-\lambda_F} \, ,
\quad
\varphi_B=\sqrt\frac{\lambda_B+1}{\lambda_B-1} \, .
\end{equation}
The resulting form of the Hamiltonian is
\begin{equation}
\tilde H_{RA} = \tilde H_0 +
\begin{pmatrix}
V_1 + \frac{1}{1-\lambda_F^2} && V_2 \cr
V_2 && - V_1 + \frac{1}{\lambda_B^2-1}
\end{pmatrix}\, ,
\label{ham-triplet-symmetric}
\end{equation}
where $\tilde H_0$ is given by
(\ref{Hamiltonian-singlet-1})--(\ref{Hamiltonian-singlet-2}) [the Hamiltonian
in the singlet sector], and
\begin{equation}
V_1=2\frac{\lambda_B \lambda_F -1}{(\lambda_B - \lambda_F)^2}\, ,
\qquad
V_2=2\frac{\sqrt{\lambda_B^2-1}\sqrt{1-\lambda_F^2}}{(\lambda_B-\lambda_F)^2}
\, .
\end{equation}
This Hamiltonian is self-conjugate with respect to the obvious
extension of the flat bilinear form (\ref{flat-form}) to two-component
vectors $\Phi=(\phi_0,\phi_1)$:
\begin{equation}
\llangle\Phi_1 | \Phi_2 \rrangle
= \int_1^\infty d\lambda_B \int_{-1}^{1} d\lambda_F\,
\Tr \Phi_1^T \Phi_2\, .
\label{flat-triplet}
\end{equation}

Proceeding along the same lines as in the singlet case, we write
\begin{equation}
\langle \str[k_{RA} Q^{RA}(0)] \str[k_{AR} Q^{AR}(\tau)] \rangle=
\langle \Psi_0 | \str[k_{AR} Q^{AR}] \exp(-H\tau) \str[k_{RA} Q^{RA}] |
\Psi_0 \rangle
\label{triplet-evolution}
\end{equation}
and use the reduction rules (\ref{magic-triplet-1})--(\ref{magic-triplet-2})
to arrive at the final result
\begin{equation}
\langle \str[k_{RA} Q^{RA}(0)] \str[k_{AR} Q^{AR}(\tau)] \rangle=
- \str (k_{RA} k_{AR}) \,
\llangle \Psi_0 \otimes u^T |  \exp(-\tilde H_{RA} \tau)
| u \otimes \Psi_0 \rrangle
\end{equation}
where
\begin{equation}
u=4 U^{-1} \begin{pmatrix} 1 \\ -1/4 \end{pmatrix}
= - \frac{1}{4} U^T \begin{pmatrix} M_1 \\ M_2 \end{pmatrix}
=\frac{1}{\lambda_B-\lambda_F}
\begin{pmatrix} \sqrt{1-\lambda_F^2} \\ -\sqrt{\lambda_B^2-1} \end{pmatrix}
\end{equation}
is an auxiliary two-component vector corresponding to the two-dimensional
space (\ref{RA-general}).

One can show\cite{Pasha-communication} that this formalism is equivalent to
that in Eqs.~(11.46) and (11.47) of Efetov's book\cite{Efetov-book}.

\section{Other symmetry classes}
\label{S:classes}

It does not appear possible to generalize the DM parameterization
to the symplectic and orthogonal random-matrix symmetry classes:
the corresponding $Q$ matrices have additional constraints
mixing the off-diagonal blocks $Q_{RA}$ and $Q_{AR}$,
and we cannot reconcile
those constraints with the asymmetric DM parameterization in terms
of $b$ and $\bbar$.

However, a generalization is possible for the symmetry classes C
and B/D. These are the ``superconducting'' symmetry classes discussed
by Altland and Zirnbauer\cite{Altland-Zirnbauer}, with the unitary-type
interlevel repulsion ($\beta=2$).

We specify our discussion below to the supersymmetric formulation of NLSM.

For the symmetry class B/D, one may impose additional constraints
on the matrices $b$ and $\bbar$:
\begin{equation}
\bbar^T k = \bbar\, , \qquad
k b^T = b ,
\label{BD-DM}
\end{equation}
where the transposition operation is understood in the supersymmetric
sense (with changing the sign of one of the Grassmann components,
see, e.g., Eq.~(2.21) of Ref.~\onlinecite{Efetov-book}), and $k$ is the
superparity operator,
\begin{equation}
k=\begin{pmatrix}1 & 0 \cr 0 & -1
\end{pmatrix}_{BF}\, ,
\end{equation}
so that for any supermatix $A$, $(A^T)^T = kAk$.
The condition (\ref{BD-DM}) used in the parameterization (\ref{Q-DM})
implies
\begin{equation}
Q = - \gamma Q^T \gamma^{-1}\, ,
\qquad \gamma=\begin{pmatrix}
0 & k \cr -1 & 0
\end{pmatrix}_{RA} ,
\end{equation}
which is the constraint on the $Q$ matrix in the B/D symmetry
class~\cite{Zirnbauer-classification}.

Similarly, the constraints
\begin{equation}
\bbar^T k = - \bbar\, , \qquad
k b^T = - b
\label{C-DM}
\end{equation}
produce the DM parameterization for the C symmetry class [the spaces of
$Q$ matrices in the classes C and B/D differ by the interchange of the
fermionic and bosonic components\cite{Zirnbauer-classification},
and thus (\ref{C-DM}) is obtained from (\ref{BD-DM}) by replacing $k$ by $-k$].

The constraints (\ref{BD-DM}) or (\ref{C-DM}) reduce the number
of independent components in $b$ and $\bbar$ by a factor of two.

Note that in the class B/D, the superspace of $Q$ matrices contains
two connected components, and the two subclasses
B and D differ by the relative sign of the contributions from those
components\cite{Ivanov-zero-modes}. This sign choice coirresponds to the
presence or absence of the zero mode in the original random-matrix theory.
In the context of the DM parameterization,
this implies that the parameterization (\ref{Q-DM}) produces
only one of the two connected components (containing the ``north pole''
$\Lambda$), and the second component (containing $k\Lambda$) may be obtained
by an appropriate non-local rotation (see, e.g.,
Ref.~\onlinecite{Ivanov-zero-modes} for details).

The only other symmetry class with $\beta=2$ is the Chiral Unitary class
(AIII in Ref.~\onlinecite{Zirnbauer-classification}). So far we were not
able to adapt the DM parameterization for this class (with the space of $Q$
matrices of the A$|$A type).

\section{Summary}
\label{S:Summary}

In this paper we discussed the Dyson-Maleev parameterization of nonlinear
sigma-models of the unitary symmetry class. Contrary to the commonly-used
Hermitian infinite-series parameterizations (\ref{Q-W}), the non-Hermitian
DM transformation (\ref{Q-DM}) parameterizes the NLSM target space
with finite-degree (cubic) polynomials.
The Jacobian of the DM parameterization is unity.
We find that the DM parameterization can be introduced only for the unitary
symmetry class and its ``relatives'' C and B/D classes characterized
by the same level-repulsion parameter $\beta=2$.

The main advantages of the DM parameterization include:
\begin{itemize}
\item Simplification of the perturbative diagrammatic expansion.
Due to the absence of higher-order interaction vertices in the DM
representation, the number of diagrams is strongly reduced compared
to the general infinite-series parameterization (\ref{Q-DM}).
Classification and identification of corresponding diagrams becomes
a rather simple combinatorial task.

\item Possibility of obtaining non-perturbative results without
resorting to Efetov-like parameterizations of the $Q$ matrix.
Instead, one can integrate directly over $b$ and $\bar b$
with flat measure, with proper constraints on the eigenvalues
of the product $b\bar b$.

\item An ``algebraic'' approach to the transfer-matrix treatment
of one-dimensional diffusive NLSM. The derivation of the transfer matrix
relies only on the symmetries of the action
and on the algebraic structure of the superspace of $Q$ matrices,
without resorting to explicit manipulations with coordinates. Furthermore,
the expressions for wave functions and correlations in terms of DM fields
manifestly specify their symmetries, which simplifies manipulation,
analysis, and interpretation of results.
\end{itemize}

We expect that the Dyson-Maleev parameterization will be useful
for perturbative treatment of various NLSM and, in particular,
of the Finkelstein's replicated sigma-model for interacting
systems \cite{Finkelstein}.
The DM transformation might also be of importance for the study
of two-dimensional localization in magnetic field since
the topological term becomes quadratic in $b$ and $\bar b$,
see Appendix \ref{A:top}.
Finally, the ``algebraic'' approach to the DM parameterization
(using symmetries and algebraic properties instead of explicit
coordinates) may provide a helpful tool for establishing relations
between the three formulations of the NLSM: supersymmetric, replica,
and Keldysh.

\acknowledgments

We thank A.~Abanov, I.~S.~Burmistrov, I.~A.~Gruzberg, P.~M.~Ostrovsky,
and A.~M.~Tsvelik for stimulating discussions.
The work by M.~A.~S. was partially supported
by RFBR grant No.\ 07-02-00976.

\appendix

\section{Finite-dimensional DM parameterizations: spaces and symmetries}
\label{Appendix:math}

In this Appendix, we describe the spaces and symmetries involved in the
DM construction in its finite-dimensional version. We restrict ourselves to the
supersymmetric formulation of the NLSM and to the case of the
unitary symmetry class without additional constraints (see the discussion of
constraints for symmetry classes C and B/D in Section \ref{S:classes}).
For the sake of generality, we take arbitrary dimensions of the bosonic
and fermionic spaces: $m_B$ and
$m_F$, respectively. This consideration will then be equally applicable to
ordinary and super spaces.

We start our construction with the two
complex linear (super)spaces $L_R$ and $L_A$ (retarded and advanced
sectors, respectively) of equal dimensions $m_B|m_F$ (we take the dimensions
equal for simplicity; the construction may be extended for $L_R$ and $L_A$ having
different dimensions).
Then we define $b$ and $\bbar$ to be
elements of the complex spaces of linear operators from $L_A$ to $L_R$ and
backwards:
\be
  b: L_A \to L_R\, , \qquad \bbar: L_R \to L_A\, .
\ee
The products
$b\bbar$ and $\bbar b$ are then linear maps
$L_R \to L_R$ and $L_A \to L_A$, respectively.

The matrix $Q$ in (\ref{Q-DM}) is a linear operator acting in the
space $L_R \oplus L_A$. In the above example of $L_R$ and $L_A$ having equal
dimensions $m_B|m_F$, the matrix $Q$ has the dimension $2m_B|2m_F$. The condition
$Q^2=1$ defines a complex manifold in this linear space of matrices, and the DM
parameterization allows us to parameterize its connected component containing
the matrix (\ref{Lambda-definition}).

On this complex manifold of $Q$-matrices, we consider the group
generated by the invertible linear transformations in $L_R$ and $L_A$
($H_R=GL(L_R)$ and $H_A=GL(L_A)$, respectively). The action of the group
$H_R\times H_A$ on the $Q$-matrix (\ref{Q-DM}) may be written in terms
of $b$ and $\bbar$
as
\begin{eqnarray}
  &&b \mapsto U_R b U_A^{-1} , \nonumber\\
  &&\bbar \mapsto U_A \bbar U_R^{-1} ,
\end{eqnarray}
with $U_R\in H_R$, $U_A\in H_A$.

Since the group $H_R\times H_A = GL( m_B | m_F ) \times GL( m_B | m_F )$
is the group of elements leaving $\Lambda$ invariant, the complex (super)space
of $Q$ matrices  (all possible rotations of $\Lambda$) may be described as
\begin{equation}
GL( 2m_B | 2m_F )/
\left[ GL( m_B | m_F ) \times GL( m_B | m_F ) \right]
\label{symmetric-superspace}
\end{equation}
This is the AIII$|$AIII symmetric space in the classification of
Zirnbauer\cite{Zirnbauer-classification}, which corresponds to
the unitary random-matrix theory (class A).

Our illustrative supersymmetric examples in Sections \ref{S:GUE} and \ref{S:Q1D}
correspond to the case $m_B=m_F=1$.
In principle, the DM parameterization
may also be used in non-supersymmetric models (with $m_B\ne m_F$):
for example, the simplest case $m_B=1$, $m_F=0$ corresponds to the original
DM parameterization (\ref{DM-spin}) for spin $S=1/2$.

Another important group is the symmetry group of the NLSM action
(\ref{S[Q]-general}). In our examples in Sections \ref{S:GUE} and \ref{S:Q1D},
the symmetry group coincides with  $H_R\times H_A$ (the stabilizer of $\Lambda$).
Then the coordinate space for singlet wave functions ($\lambda_B$, $\lambda_F$) may
be mathematically described as the double quotient
\begin{equation}
\left[ GL( m_B | m_F ) \times GL( m_B | m_F ) \right]
\backslash GL( 2m_B | 2m_F )/
\left[ GL( m_B | m_F ) \times GL( m_B | m_F ) \right] .
\label{double-quotient}
\end{equation}

\section{Topological term in the DM representation}
\label{A:top}

The sigma-model action describing the low-energy dynamics
of the two-dimensional electron system subject to a strong
perpendicular magnetic field is given by \cite{Pruisken1984}
\be
\label{S-top}
  S
  =
  - \frac{\sigma_{xx}}{8} \int d{\bf r}
    \tr (\nabla Q)^2
  + \frac{\sigma_{xy}}{8} \int d{\bf r}
    \tr \eps_{\mu\nu} Q \nabla_\mu Q \nabla_\nu Q ,
\ee
where $\sigma_{xx}$ and $\sigma_{xy}$ are the mean-field
longitudinal and Hall conductances respectively,
and $\eps_{\mu\nu}$ is the antisymmetric tensor.
The last term in the action (\ref{S-top}) is usually
referred to as the topological term.

In the Dyson-Maleev parameterization (\ref{Q-DM}), the action reads
\be
  S
  =
  - \frac{\sigma_{xx}}{4} \int d{\bf r}
    \tr \nabla b \nabla \bar b
  - \frac{\sigma_{xx}}{16} \int d{\bf r}
    \tr \nabla b \bar b \nabla b \bar b
  + \frac{\sigma_{xy}}{4} \int d{\bf r}
    \tr \eps_{\mu\nu} \nabla_\mu b \nabla_\nu \bar b .
\ee
Note that the topological term becomes {\em quadratic}\/ in $b$ and $\bar b$,
whereas all nonlinearity originates from the first term
in the action~(\ref{S-top}).

\end{document}